\documentstyle[epsfig]{mn}

\title[]
{The effects of population III stars on the chemical and photometrical 
evolution of ellipticals}

\author[F. Matteucci \& A. Pipino]{Francesca Matteucci and Antonio Pipino
\\
Dipartimento di Astronomia, Universit\`a di Trieste,
    Via G.B. Tiepolo 11, I-34127, Trieste, Italy}

\date{Accepted,
      Received }

\begin{document}
\maketitle

\begin{abstract}
We have studied the effects of an hypothetical initial generation containing
very massive stars ($M > 100 M_{\odot}$, pair-creation SNe) on the 
chemical and photometric evolution of elliptical galaxies.
To this purpose, we have computed the evolution of a typical elliptical 
galaxy with luminous mass of the order of 
$10^{11}M_{\odot}$ and adopted chemical evolution models already 
tested to reproduce the main features of ellipticals.
We have tested several sets of yields for very massive zero-metallicity stars:
these stars should produce quite different amounts of heavy elements 
than lower mass stars.
We found that the effects of population III stars on the chemical 
enrichment is negligible if 
only one or two generations of such stars occurred, whereas they produce 
quite different results from the standard models if they continuously formed 
for a period not shorter than 0.1 Gyr.
In this case, the results are at variance with the main 
observational constraints of ellipticals such 
as the average [$<\alpha/Fe>_{*}$] ratio in stars and the integrated 
colors.
Therefore, we conclude that if population III stars ever existed they must 
have been present for a very short 
period of time and their effects on the following evolution of the parent 
galaxy must have been negligible. 
This effect is minimum if a more realistic model with initial infall of gas 
rather than the classic monolithic model is adopted. Ultimately, we 
conclude that there is no need to invoke a generation of very massive stars 
in ellipticals to explain their chemical and photometric properties.
\end{abstract}


\section{Introduction}
The first stars formed out of gas with primordial chemical composition, 
namely without metals.
Therefore, if the stellar mass distribution has not changed drastically 
from early times to now we would expect to find turn-off mass stars showing 
no metals in their atmospheres, but so far they have not been detected.
However, in the past years stars with extremely low metal content down to 
[Fe/H]=-5.3 (Christlieb et al. 2002;2004) and residing in the Galactic halo
have been found and it is likely that more of these stars 
will be observed in the 
future thanks to the constant improvement of the astronomical techniques
and to the larger and larger stellar data samples.

From a theoretical point of view we expect very few stars with 
extremely low metal content since a few massive supernovae exploding at 
early times are sufficient to pollute  the interstellar medium at 
levels far from zero. For example, in Chiappini et al. (2003) model for 
the evolution of the Milky Way it is found that to enrich the galactic halo 
up to [Fe/H]=-5.0, with the most massive stars being those of $100M_{\odot}$,
it takes only between 3 and $4 \cdot 10^{6}$ years, with the consequence of having a very small fraction (of the order of $10^{-5}$)
of stars formed with metallicity below
[Fe/H]=-5.0. 
Prantzos (2003) also reached similar conclusions.
In the Chiappini et al. model 
the halo is assumed to be well mixed at any time, an assumption which seem 
to be reasonable given the lack of spread observed in the abundances of halo 
stars 
down to [Fe/H]=-4.0 (Cayrel et al. 2004: Fran\c cois et al. 2004).
However, due to the lack of observed zero-metal stars, the idea of 
a primordial stellar generation made only of very massive stars, the 
so-called 
population III stars (Bond, 1981; Cayrel 1986; Carr 1987, 1994), 
has fascinated astronomers for many years.
Recently, the WMAP experiment has observed the large- angle polarization 
anisotropy of the cosmic microwave background (CMB), thus constraining 
the total production of ionizing photons from the first stars 
(e.g. Cen et al. 2003 a,b; Ciardi et al. 2003), and suggesting a 
substantial early activity of massive star formation at redshifts $z \ge 15$.
Moreover, the recent realization that the intergalactic medium seems to be 
enriched in metals even at high redshift, as indicated by the presence of 
metals in the Ly$\alpha$ forest at $z \sim$ 4-5 (e.g. Songaila 2001, 
Schaye et al. 2003) has also suggested the possibility 
of a population III generation of stars.

From the point of view of stellar evolution, the evolution and 
nucleosynthesis of zero-metal massive and very massive stars has been 
computed since the early eighties 
(e.g. Carr et al. 1982; Ober et al. 1983; 
El-Eid et al. 1983) and has continued until recently 
(e.g. Woosley \& Weaver, 1995; Heger \& Woosley 2002; Umeda \& Nomoto, 2002;
Heger et al. 2003; Chieffi \& Limongi 2004).
Even studies of the evolution of intermediate mass (e.g. Chieffi et al. 2001;
Siess et al. 2002; Iwamoto et al. 2004) and low mass stars 
(D'Antona 1982) with zero metallicity have appeared in the 
literature in the past years.

\par
The most important question about population III stars is to decide 
how massive they were. 
Hydrodynamical simulations of the collapse and fragmentation of the 
primordial gas clouds suggest that the very first stars should have been 
indeed with masses larger than $100 M_{\odot}$ (Bromm 1999, 2002; 
Abell et al. 2000;2002).
Some of the stellar evolution studies mentioned before computed the 
evolution and nucleosynthesis of stars with main sequence masses larger than 
100$M_{\odot}$ (Ober et al. 1983; El Eid et al. 1983; Umeda \& Nomoto, 2002;
Heger \& Woosley, 2002). The interesting aspect of these very massive stars is that they should die as pair-creation supernovae and leave no remnants. 
These stars, in fact, are first characterized by the nuclearly energized pulsational
instability during the phases of core H- and He-burning, which can produce substantial mass loss.
Later on, a pair creation instability sets in and induces core collapse, explosive O-burning and subsequent SN explosion which disrupts completely the star.
For masses larger than 200$M_{\odot}$ it can instead occur that they 
implode as black holes and do not contribute to the chemical enrichment.
\par
The effect of population III stars on the chemical evolution of galaxies 
has been studied a long ago (e.g. Carr et al. 1982; 
Chiosi \& Matteucci 1982) and it was suggested that they could explain 
the trend of [O/Fe] versus 
[Fe/H] observed in the Milky Way stars. However, at that time, detailed 
calculations of the effect of type Ia SNe in explaining that trend had 
not yet 
appeared (Greggio \& Renzini, 1983; Matteucci \& Greggio 1986).
Nowadays, we interpret the run of [O/Fe] vs. [Fe/H] as due to the 
time-delay between the production of oxygen by type II SNe and
Fe by type Ia SNe.
However, in spite of the better  knowledge of chemical evolution that we 
have now relative to then, we think that it is still interesting to study the 
effects of population III stars on the chemical evolution, in order to impose 
constraints on the number and masses of these hypothetical first stars.

Recent studies dealing with the effects of population III stars on the 
chemical enrichment have concluded either that stars with masses between 
1-40$M_{\odot}$ must have existed together with pair creation SNe to 
explain the abundance pattern observed in high redshift QSOs 
(Venkatesan et al. 2004), or that 
the nucleosynthesis in pair creation SNe cannot reproduce the abundance 
patterns of halo stars (Umeda \& Nomoto, 2002).
However, in the first paper only an instantaneous starburst was considered 
and in the second only a simple comparison between stellar
production ratios and abundances was performed.
\par
In this paper we have explored the consequences of introducing one or 
more generations made either only of very massive stars 
($M>$100$M_{\odot}$) or made of all stars plus the 
very massive ones (0.1-260$M_{\odot}$)
in detailed chemical and photometric evolution 
models for a typical elliptical galaxy. 
The chemical and photometric models adopted for the elliptical galaxy is that 
of Pipino \& Matteucci (2004) which best reproduces the main properties of 
ellipticals. This model belongs to the category of supernova-driven wind 
models and assumes that in ellipticals there is an early strong burst of 
star formation followed by the development of a galactic wind. After the 
wind starts, and
this occurs always at times shorter than 1 Gyr, star formation is 
inhibited until the present time and the galaxy evolves passively.
The detailed nucleosynthesis from SNe II and Ia is taken into account.
In this paper we add the nucleosynthesis from early pair creation supernovae.
The paper is organized as follows: in section 2 we describe the model
and the prescriptions adopted for population III stars, in section 3 the 
results are shown and in section 4 some conclusions are drawn.


\section{The model}

The chemical code adopted here is described in full detail 
in Pipino \& Matteucci (2004, PM04 hereafter). In particular, 
we refer to PM04 Model II  input parameters as our \emph{standard model}.
This model is characterized by:
Salpeter (1955) IMF, Thielemann et al (1996) yields for massive stars,
Nomoto et al (1997) yields for type Ia SNe and 
van den Hoek \& Groenewegen (1997) yields for low-
and intermediate-mass stars. 
We limit our analysis to a $10^{11}M_{\odot}$ galaxy, but it is 
easy to show that
the results are more general. 

The model assumes that the galaxy assembles by merging of gaseous lumps 
(infall) on 
a short timescale and suffers a strong starburst which injects into the 
interstellar medium (ISM) a large amount of energy which triggers
galactic winds.
After the development of the wind, the star formation is assumed to stop 
and the galaxy evolves passively with continuous mass loss.

We recall here that the assumed star formation efficieny 
is $\nu =10 \rm Gyr^{-1}$,
while the infall timescale is $\tau =0.4 \rm Gyr$ in the galactic core 
and $\tau =0.01 \rm Gyr$
at $1R_{eff}$, respectively. 
A dark matter halo ten times more massive than the luminous mass
and with a ratio between the effective radius and the radius of the dark 
matter core ${R_{eff} \over R_d}=0.1$, are adopted.
These values were chosen in PM04 in order to reproduce
the majority of the chemical and photometric properties of ellipticals.

The photometric code is the one of Jimenez et al. (1998) 
(see details in PM04).

\smallskip
We run several models by changing the characteristics of the population 
III stars, but keeping a standard Salpeter (1955) IMF, with x=1.35 over the 
whole mass range. 

In particular, we considered the following cases:
\begin{itemize}
\item[Case a:] It is based on PM04 Model II. 
In particular, 
the reference cases \emph{a1c}, \emph{a1o} 
(where \emph{c} stands for \emph{core} and \emph{o} for 
\emph{outskirts}, respectively) are taken
from PM04, their Table 5. 
In the other $a$ models,
we allow for the simultaneous presence of both pair-creation SNe and 
and stars of all masses (down to 0.1$M_{\odot}$). 
Then we assume that after 
a time $\Delta t_{PopIII}$, in which the gas
has attained a threshold metallicity 
$Z_{thr} \sim 10^{-3}Z_{\odot}$ (Umeda \& Nomoto, 2002; Bromm et al. 2001),
only stars with masses $< 100 M_{\odot}$ are formed.
However, it is worth noting that the results do not change if we adopt a threshold $Z_{thr}= 5 \cdot 10^{-2}Z_{\odot}$. Values as low as $Z_{thr}= 
10^{-5 \pm 1}Z_{\odot}$, as suggested by Schneider et al. (2002) are obtained 
already with one single generation of very massive pop III stars, analogously to what happens in the Milky Way models.

Cases in which stars in the mass range $40 - 100 M_{\odot}$ end their 
life as black holes,
without contributing to the pollution of the ISM, were run for comparative 
purposes (labelled as \emph{BH)}.

\item[Case b:] As \emph{a}, but without initial infall, namely they are 
based on 
the closed-box (CB) monolithic scenario. Therefore, the model 
presents a star formation rate peaked
at the beginning of the galactic evolution 
(i.e. when pop III stars are thought to form).

\item[Case c:] Here we assume a strongly bimodal star formation history: in the early 
stages, only very massive stars ($M >100 M_{\odot}$) ending as
pair creation SNe are allowed to form. After a delay $\Delta t_{PopIII}$,
in which the star formation is possibly halted by an early galactic wind, 
the stars form  with a standard
Salpeter IMF on a range of 0.1-100$M_{\odot}$. 
The infall and star formation timescales are the same as those of Model II
of PM04. 
The time delay, which is either 0.01 or 0.1 Gyr, is considered here
as a free parameter in order to see how
its variation could alter the photochemical properties of the galaxy. 
\end{itemize}
In Table 1 we present the characteristics of the models:
in particular, in the first column are shown the model names, in the
second column is indicated the reference model we used (II stays for 
Model II of PM04 and CB for closed-box), in column 3 is indicated the 
considered spatial range in terms of galactic effective radius and 
in column 4 is shown 
whether we have considered stars in the range 40-100$M_{\odot}$ 
to be active in enriching the ISM or to collapse into black holes.

\subsection{Population III yields}
Different sets of pair creation SNe yields were taken into account, in 
particular:
\begin{itemize}
\item From Heger \& Woosley (2002) (hereafter HW02).
These yields are calculated for a zero metallicity chemical composition 
and for stars with He-core masses in the range 64-133$M_{\odot}$ 
corresponding to main sequence masses in the range 140-260$M_{\odot}$. 
For He-core masses larger than 133$M_{\odot}$, these authors find that a 
black hole is formed and no nucleosynthesis is produced. They also find that 
the same situation is likely
to occur in stars with main sequence masses in the range 25-140$M_{\odot}$
and primordial chemical composition.
\item From Umeda \& Nomoto (2002) (hereafter UN). These authors computed 
the nucleosynthesis of zero metallicity stars of main sequence masses in 
the range 13-30$M_{\odot}$ and in the range 130-300$M_{\odot}$. 
They compared the abundance ratios of halo stars with the 
production ratios 
in pair-creation SNe and concluded that they are not in agreement.
\par
We did not
consider the more recent pop III star yields by Chieffi \& Limongi (2004) 
since they do not contain stars above $35 M_{\odot}$.

\item From Ober et al. (1983) (hereafter OFE83). These authors computed the 
yields for zero metallicity stars with main sequence masses in the range 
80-500$M_{\odot}$. They predicted that these objects should produce mainly 
oxygen and strongly underproduce nitrogen and Fe-peak nuclei.
We are aware that these calculations are old and could have been superseeded 
by the previously mentioned calculations, but we think that it can be 
interesting to compare their results with the new ones, especially because 
the OFE83 yields look quite different from the others.

\end{itemize}
For the stars between 0.8 (turn-off mass) and 100$M_{\odot}$ of 
any chemical composition we adopted the results 
of Thielemann et al. (1996) which are relative to the solar chemical 
composition for stars with masses larger
than 10$M_{\odot}$, whereas for the low and intermediate mass stars we kept 
the yields of van den Hoek and Groenewegen (1997) which are computed for 
different metal contents, including the zero-metallicity composition.
This was done for the sake of simplicity, in order to be able to compare the 
results with those of PM04.

In Table 2 are summarized the main properties of the models for what 
concerns the adopted nucleosynthesis.
In particular, the model names are given in the first column as in Table 1, 
while
the second column indicates whether a population III has been considered or 
not, the third column shows the adopted mass range for the pop III stars.
In column 4 are indicated the sources of the adopted yields for the pop III 
stars, in column 5 the time during which we allow the pop III stars to form 
and in column 6 is indicated whether the pop III stars are able to trigger a 
very early galactic wind. 

\begin{table*}
\centering
\begin{minipage}{120mm}
\begin{flushleft}
\caption[]{Model definition: formation process and galactic shell considered}
\begin{tabular}{llll}
\hline
\hline

Model  & based on &zone & 40-100 \\
\hline
a1c&II & 0-0.1 $R_{eff}$ & yes \\
a1o&II & 0.9-1 $R_{eff}$ & yes \\
a1BH&II & 0-0.1 $R_{eff}$ &no\\
a2c&II& 0-0.1 $R_{eff}$ & yes\\
a2o&II & 0.9-1 $R_{eff}$ & yes\\
a2BH&II& 0-0.1 $R_{eff}$ &no\\
a3c&II& 0-0.1 $R_{eff}$ & yes\\
a3o&II& 0.9-1 $R_{eff}$ & yes\\
a4&II& 0-0.1 $R_{eff}$ & yes\\
\hline
No infall\\
\hline
b1&CB & 0-0.1 $R_{eff}$ & yes\\
b1BH&CB & 0-0.1 $R_{eff}$ & no\\
b2&CB& 0-0.1 $R_{eff}$ & yes\\
b3&CB& 0-0.1 $R_{eff}$ & yes\\
b4&CB& 0-0.1 $R_{eff}$ & yes\\
\hline
Strongly bimodal\\
\hline
c1&II & 0-0.1 $R_{eff}$ & yes\\
c2&II & 0-0.1 $R_{eff}$ & yes\\
c3&II & 0-0.1 $R_{eff}$ & yes\\
c4&II & 0-0.1 $R_{eff}$ & yes\\
c5&II & 0-0.1 $R_{eff}$ & yes\\
c6&II & 0-0.1 $R_{eff}$ & yes\\
c7&II & 0-0.1 $R_{eff}$ & yes\\
c8&II & 0-0.1 $R_{eff}$ & yes\\
\hline
\end{tabular}
\end{flushleft}
\end{minipage}
\label{tab-mod1}
\end{table*}

\begin{table*}
\centering
\begin{minipage}{120mm}
\begin{flushleft}
\caption[]{Model defintion: PopIII properties}
\begin{tabular}{llllll}
\hline
\hline

Model  & PopIII & Mass range (PopIII) &yields (PopIII) & $\Delta t_{PopIII}$ &early w.\\
       &        &   ($M_{\odot}$)     &                &   (Gyr)             &        \\
\hline
a1c& no & - & - & - & no\\
a1o& no & - & - & - & no\\
a1BH& no & - & - & &no\\
a2c& yes & 0.1-260&HW02 & 0.004 &no\\
a2o& yes & 0.1-260&HW02 & 0.004&no\\
a2BH& yes & 0.1-260&HW02 & 0.004&no\\
a3c& yes & 0.1-270&UN & 0.004 &no\\
a3o& yes & 0.1-270&UN & 0.004 &no\\
a4&yes &0.1-220&OFE83 & 0.006 &no\\
\hline
No infall\\
\hline
b1& no & - & - & - &no\\
b1BH&no & - & - & - &no\\
b2&yes & 0.1-260&HW02 & 0.004 &no\\
b3&yes & 0.1-270&UN & 0.004 &no\\
b4&yes & 0.1-220&OFE83 & 0.006 &no\\
\hline
Strongly bimodal\\
\hline
c1&yes & 140-260&HW02 & 0.001 &no\\
c2&yes & 140-260&HW02 & 0.010 &no\\
c3&yes & 140-260&HW02 & 0.100 &no\\
c4&yes & 140-260&HW02 & 0.100 &yes\\
c5&yes & 140-270&UN & 0.010 &no\\
c6&yes & 140-270&UN & 0.100 &no\\
c7&yes & 108-220&OFE83 & 0.010 &no\\
c8&yes & 108-220&OFE83 & 0.100 &no\\
\hline
\end{tabular}
\end{flushleft}

\end{minipage}
\label{tab-mod2}
\end{table*}

\section{Results and discussion}

In Table 3 and in Figures 1-6  we present the results of the 
above described calculations.
In particular, in the second colunm of Table 3 is shown the predicted 
[$<Mg/Fe>_{*}$] ratio in the dominant stellar population of 
the studied galaxy. 
This value is obtained by performing a mass weighted average,
as described in PM04.
In column 3 and 4 are shown the mass averaged [$<Fe/H>_{*}$] and 
[$<Mg/H>_{*}$], respectively.
Then in columns 5, 6 and 7 are shown the predicted colors U-V, V-K and J-K
(only for the central galactic regions), 
respectively.

\begin{table*}
\centering

\begin{minipage}{120mm}
\begin{flushleft}

\caption[]{Model predictions}
\begin{tabular}{lccclll}
\hline
\hline

Model  &$[<Mg/Fe>_*]$&  $[<Fe/H>_*]$&$[<Mg/H>_*]$	&U-V&	V-K&	J-K\\
\hline
a1c&0.206&0.116&0.07&1.39&3.10&1.03\\
a1o&0.532&-0.145&0.02&-&-&-\\
a1BH&-0.03&0.06&-0.218&1.17&2.75&0.91\\
a2c&0.213&0.08&0.06&1.38&3.07&1.02\\
a2o&0.415&-0.09&0.04&-&-&-\\
a2BH&-0.04&0.06&-0.217&1.17&2.75&0.90\\
a3c&0.213&0.07&0.06&1.42&3.13&1.04\\
a3o&0.458&-0.08&0.04&-&-&-\\
a4&1.03&0.07&0.05&1.74&3.60&1.18\\
\hline
No infall\\
\hline
b1&0.507&-0.06&0.04&0.76&2.73&0.93\\
b1BH&0.300&-0.09&-0.247&1.02&2.69&0.89\\
b2&0.342&0.116&0.231&1.81&3.76&1.23\\
b3&0.382&0.100&0.209&1.82&3.76&1.24\\
b4&4.66&-0.06&0.06&1.18&3.00&1.00\\
\hline
Strongly bimodal\\
\hline
c1&0.215&0.09&0.06&1.39&3.11&1.03\\
c2&0.156&0.10&0.08& 1.43&3.14&1.04\\
c3&0.119&0.73&0.83& 2.17&3.97&1.27\\
c4&0.166&1.50&1.64&2.36&3.91&1.26\\
c5&0.161&0.10&0.08&1.43&3.13&1.04\\
c6&0.122&0.66&0.77& 2.21&3.99&2.12\\
c7&1.070&0.09&0.09&1.31&2.98&0.99\\
c8&2.56&0.01&1.02& 2.18&3.98&1.27\\
\hline
\end{tabular}
\end{flushleft}
\end{minipage}
\label{tab-res}
\end{table*}

\begin{figure}
\epsfig{file=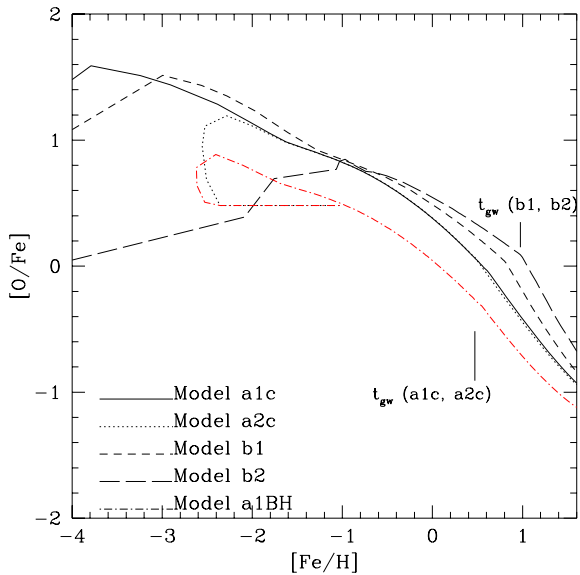, height=8cm, width=8cm}
\caption{The [O/Fe] vs. [Fe/H] relations in the gas of a typical elliptical 
galaxy as predicted by various models. The time for the occurrence of a 
galactic wind is also marked on the curves. The time for model \emph{a1BH} is the 
same as for model  \emph{a1c}.}
\label{fig1}
\end{figure}

\begin{figure}
\epsfig{file=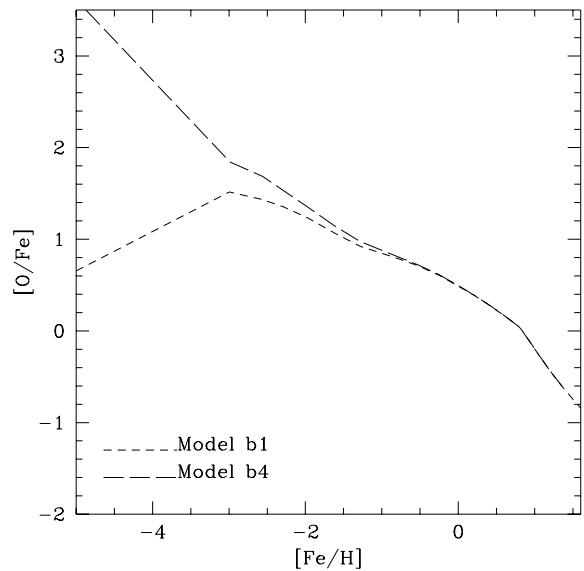, height=8cm, width=8cm}
\caption{The [O/Fe] vs. [Fe/H] relations in the gas of a typical elliptical 
galaxy as predicted by models \emph{b1} and \emph{b4} (nucleosynthesis from OFE83). The time for the occurrence of a 
galactic wind is not marked since it corresponds to the same as in models 
\emph{b1} and \emph{b2}, already indicated in Figure 1.}
\label{fig2}
\end{figure}

\begin{figure}
\epsfig{file=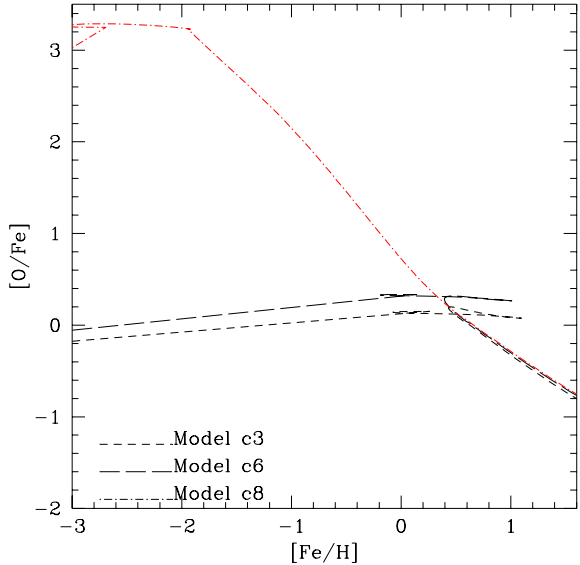, height=8cm, width=8cm}
\caption{The [O/Fe] vs. [Fe/H] relations in the gas of a typical elliptical 
galaxy as predicted by \emph{c3}, \emph{c6} and \emph{c8} models (strongly bimodal cases with 
different nucleosynthesis prescriptions for pair-creation SNe). 
The time for the occurrence of a 
galactic wind is not marked on the curves since it corresponds to the same 
one as in models  \emph{a1c} and  \emph{2c}.} 
\label{fig3}
\end{figure}

\begin{figure}
\epsfig{file=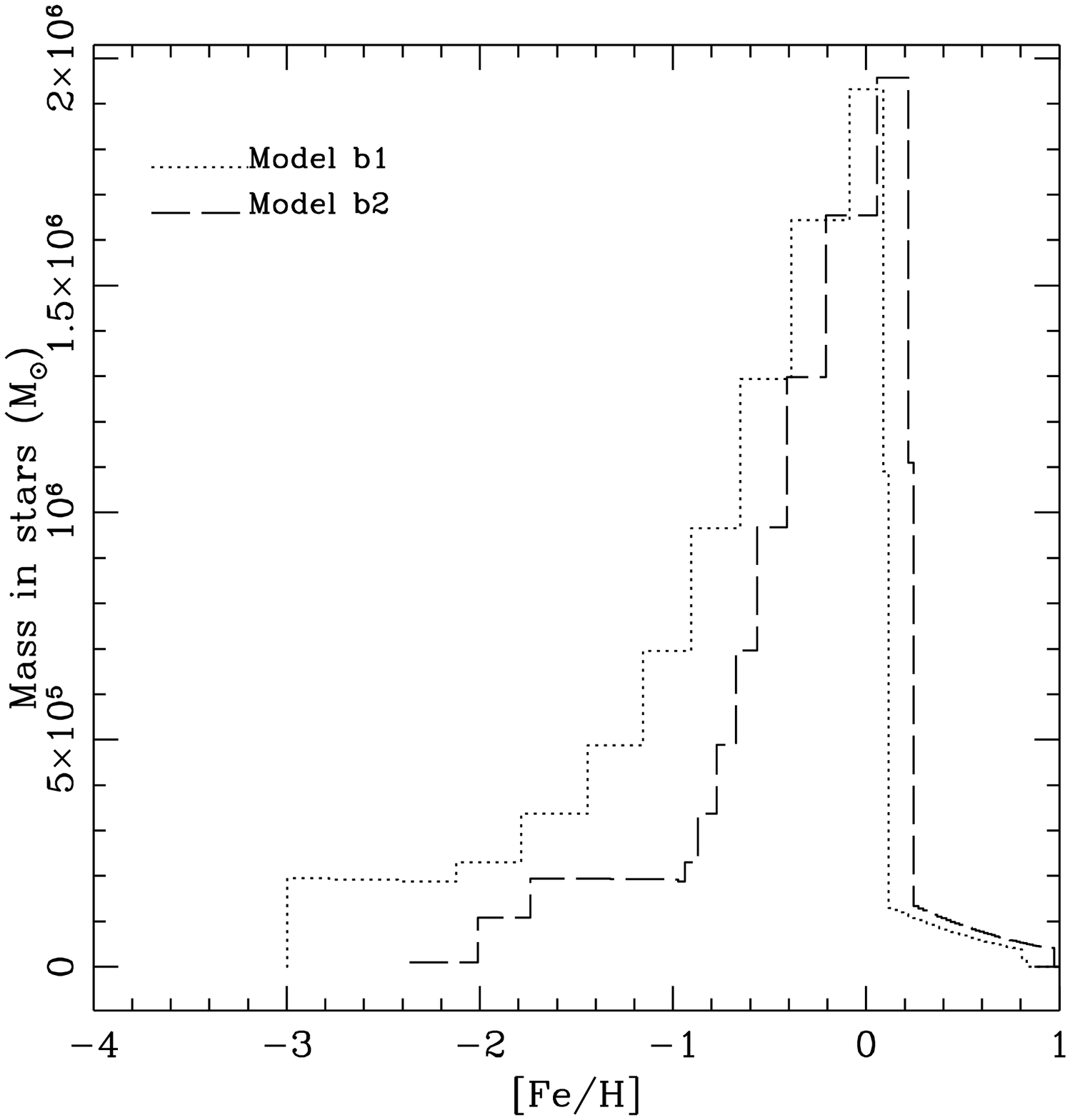, height=8cm, width=8cm}
\caption{Predicted distribution of stars as a function of metallicity 
([Fe/H]) for models \emph{b1} and \emph{b2}. They differ only for the existence of 
a population ofprimordial stars including pair-creation SNe.}
\label{fig4}
\end{figure}

The cases  \emph{a1c} and  \emph{a1o} represent the best model of PM04 and well fit the 
central values of the metallicity indices $Mg_2$ and $<Fe>$ as well as 
the central values of the colors.

The models including the nucleosynthesis from pair creation supernovae by HW02
produce results which differ only slightly from the results of the standard
case. In Figure 1 we show the predicted [O/Fe] versus [Fe/H] diagram 
(oxygen is a typical $\alpha$-element) for 
several models: as one can see, model  \emph{a2c}, with the yields by HW02, 
predicts exactly the
same behaviour of the [O/Fe] ratio as model \emph{a1c} (standard case withot 
pair-creation SNe) except for the very early phases. The case \emph{a2c} with popIII 
pair- creation SNe and HW02 yields, in fact, starts with a quite lower [O/Fe] 
ratio, relative to the standard case, due to the 
fact that pair creation SNe favor the production of 
Fe (at variance with the results of OFE83). This ratio stays 
constant while the [Fe/H] decreases and then increases to reach the value of 
the standard case when the pair-creation supernovae 
disappear. 
This inversion in case \emph{a2c} is due to the presence of infall of 
material of primordial chemical composition. In fact,  when the first very 
massive stars die, the [Fe/H] in the ISM jumps immediately at the value of 
[Fe/H]=-1.0 but soon this value decreases due to the infalling gas.
Model \emph{b2} is the equivalent of the standard model without infall (CB). 
In this case, the model predicts a lower [O/Fe] ratio at the beginning 
which increases 
later on, but no inversion in the [Fe/H].

In figure 2 we show the same plot of [O/Fe] vs. [Fe/H] for the models \emph{b4} and 
b1 (shown again for comparison). Model \emph{b4} differs from model \emph{b1} only for the nucleosynthesis
in pair-creation SNe, which is from OFE83. 
One can immediately notice the 
large difference in the predictions of the two models: model \emph{b4} predicts a 
very high oxygen overabundance relative to Fe in the very early phases, due 
to the lack of Fe-peak elements in the OFE83 yields. 
In Figure 3 we show the [O/Fe] vs. [Fe/H] for the strongly bimodal star 
formation cases (only very massive stars in the early phases) 
lasting for 0.1 Gyr.
Models \emph{c3} (yields HW02) and \emph{c6} (yields UN) show a 
rather constant [O/Fe] 
ratio over the whole [Fe/H] range. This is due to the fact that the very 
massive pop III stars, in the most recent nucleosynthesis calculations, 
produce an almost solar O/Fe ratio and this will predominate also in the 
subsequent evolution.  On the other hand, model \emph{c8} shows a very high oxygen 
overabundance relative to Fe, again due to the lack of Fe-peak elements 
in the yields of OFE83.
The other \emph{c} models not included in the Figure, where the pop III stars 
form only for a very short time interval (0.01 Gyr) do not produce 
noticeable 
differences in the results relative to the standard $a1$ model.

\begin{figure*}
\epsfig{file=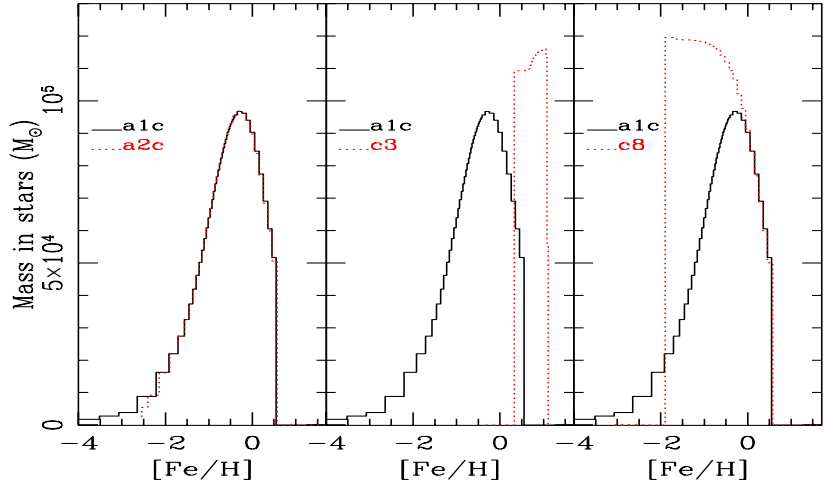, height=8cm, width=12cm}
\caption{The same as figure 4 but for models  \emph{a1c},  \emph{a2c}, 
 \emph{c3} and  \emph{c8}. The 
difference between the case with very massive primordial stars ( \emph{a2c})
and 
the standard one ( \emph{a1c}, 
no very massive stars) is negligible. On the other hand, the 
difference between case  \emph{c3} (only very massive pop III stars, nucleosynthesis 
from HW02) and  \emph{a1c}  is quite noticeable, due to the fact that pop III stars 
produce immediately a huge metal enrichment. Also model  \emph{c8} (only very massive pop III stars, nucleosynthesis from OFE83) produces a very different  distribution relative to  \emph{a1c}.}
\label{fig5}
\end{figure*}

While the abundances in Figures 1, 2 and 3 refer to the gas, 
in Figures 4 and 5  
we show the predicted distributions of stars as functions 
of [Fe/H] for models  \emph{a1c} -  \emph{a2c},  \emph{b1}-\emph{b2}, 
 \emph{c3}- \emph{a1c}.

In models  \emph{a1c}- \emph{a2c} the predicted stellar distributions are almost 
indistinguishable except for the absence of stars with [Fe/H]$<-3.0$ 
in the case with 
pop III stars. The reason for this resides in the fact that the pop III phase 
is very short and at the same time the star formation rate is small at early 
stages when there is little gas. In the closed-box cases 
(models \emph{b1} and \emph{b2}) the difference 
is more noticeable since at the 
beginning the star formation is quite high. In both models, in fact, 
no stars with metallicity lower than -3.0 and -2.0, respectively, are 
predicted (see Figure 4).
The distribution of stars with metallicity, in turn, influences the 
calculation of the average $[<Mg/H>_{*}]$ and $[<Fe/H>_{*}]$ shown 
in table  3.
For the  \emph{a1c} and  \emph{a2c} models clearly there is very little 
difference in these 
average values, whereas for \emph{b1} and \emph{b2} models the 
average $[<Mg/Fe>_{*}]$ varies 
from +0.507 to +0.32. Therefore, we conclude that the effect of pop III 
pair-creation SNe cannot be detected if ellipticals formed out of initial merging 
of primordial gas (infall cases).

\begin{figure*}
\epsfig{file=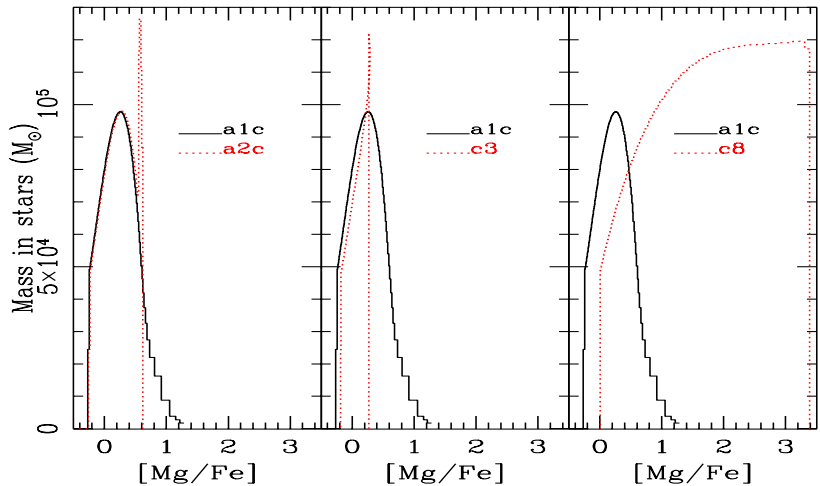, height=8cm, width=12cm}
\caption{Predicted distribution of stars as a function of the [Mg/Fe] 
ratio for models  \emph{a1c},  \emph{a2c},   \emph{c3} and \emph{c8}. 
As one can see, model \emph{c8} can 
clearly be discarded since it predicts that the majority of stars should 
have [Mg/Fe]$ >1$, at variance with observational data. Model  \emph{c3} 
shows  
a very narrow distribution, implying that most of the stars should have 
the same [Mg/Fe], and this is also unrealistic.}
\label{fig6}
\end{figure*}

The cases  \emph{a1} and \emph{a2} labelled BH and b1 labelled BH (no chemical enrichment 
from stars in the range 40-100$M_{\odot}$ in pop III and the 
following populations), predict in general
lower $[<Mg/Fe>_{*}]$, $[<Mg/H>_{*}]$ and $[<Fe/H>_{*}]$ 
than the corresponding cases where chemical enrichment from those stars is 
taken into account (see Table 3). 
This effect is not negligible as shown for the Milky Way by 
De Donder \& Vanbeveren (2003).
However, if we consider this possibility only in pop III stars, then 
only negligible effects are produced on the results, and for this 
reason we did not show this case.
\par
The cases \emph{b2}, \emph{b3}  and \emph{b4}, differing for the pop III yields, indicate that the 
yields of HW02 and UN are similar, whereas the old yields by 
OFE83 produce quite different results especially in 
the predicted 
$[<Mg/Fe>_{*}]$ which raises up to 4.66 as opposed to $\sim 0.3$ in 
the other two cases. This is due to the fact that the yields of 
OFE83 do
not contain Fe-peak elements for pair-creation SNe whereas the other yields 
do. Clearly this case should be discarded because it is at odd with 
observations
which indicate a $[<Mg/Fe>_{*}]$ $\sim$0.2-0.3 (e.g. Thomas et al. 2002; 
PM04) in ellipticals.
\par

The strongly bimodal cases (\emph{c3} and \emph{c8}), where 
the first stellar generations are made 
only of pair-creation SNe, show that, 
if the phase during which only very massive stars form, is quite short 
($< 10^{7}$ years), nothing changes relative to the standard case. 
If instead, the duration of the massive stars phase is as long as 0.1 Gyr, 
then we obtain a too high metallicity for the next stellar generations 
(see $[<Mg/H>_{*}]$  and $[<Fe/H>_{*}]$ in model \emph{c3}), with the consequence of 
obtaining too high metallicity indices and too red integrated colors.
We tried to adjust the star formation and infall parameters but could 
not find any acceptable solution. Therefore, such a case should be discarded.
The failure of bimodal star formation in elliptical 
galaxies had already 
been discussed by Gibson (1996). In particular, he showed that this assumption leads to a metallicity -luminosity relation at variance with observations.
It is worth noting that in the case \emph{c4} we adopted a 
blast wave energy for 
pair-creation SNe of $5 \cdot 10^{52}$ erg per SN, whereas in all the other 
cases we assumed the canonical value of $10^{51}$ erg.
In this case a very early ($t \sim 2 \cdot 10^{6}$ years)
galactic wind develops due to the energy injected by pair-creation SNe,
and it does not destroy the galaxy provided that 
only a small fraction of gas is lost, then star formation starts again 
with a normal IMF when the gas has cooled down (at $\sim 0.1$ Gyr).  
After this point,  the evolution is the same as in the standard case with 
galactic winds produced by normal type II and Ia SNe.
Clearly, this is a rather arbitrary situation since there is the possibility 
of destroying the whole object but we made these assumptions only for the 
sake of testing the chemical effects of a generation of very massive 
stars.
Always in case \emph{c4}, from the point of view of the enrichment of the intracluster medium (ICM), it is worth noting that the pop III chemical composition 
does not dominate over the ejecta of the late galactic winds produced by 
normal type II and Ia SNe, and this is because the galaxy can survive, 
in spite of the energy injected  by the pop III
stars, only if it looses a small fraction of gas during the very early wind.
Anyway, also in case c4 the predicted metallicities are too high and 
therefore it should be discarded.
 
Finally, in Figure 6 we present the stellar distributions 
as functions of 
the [Mg/Fe] ratio for models \emph{a1c}, \emph{a2c}, \emph{c3} and \emph{c8}. 
As one can see, model \emph{c8} can 
clearly be discarded since it predicts that the majority of stars should 
have  $[<Mg/Fe>_{*}] >1$, at variance with observational data. 
Model \emph{c3} shows  
a very narrow distribution, implying that most of the stars should have the same 
$[<Mg/Fe>_{*}]$, and this is also unrealistic. Model \emph{a2c}, as expected 
has a stellar distribution as a function of [Mg/Fe] more similar to that of 
model \emph{a1c}, although it shows a higher peak and is 
truncated at the highest values of [Mg/Fe], at 
variance with the \emph{a1c} case. 
Therefore, in general, models with pop III stars seem to worsen the agreement with observations with exception of the predicted gradient of  $[<Mg/Fe>_{*}]$
in ellipticals.
An interesting fact is that model \emph{a2} predicts a shallower $[<Mg/Fe>_{*}]$ radial gradient
with respect to the PM04 standard case, 
more in line with recent observations
claiming no variation of $[<Mg/Fe>_{*}]$ with radius (Mehlert et al. 2003).
In fact, in the galactic outskirts, where the star formation is limited to a
shorter period and it is less intense than in the center,
the effect of the pop III stars is more evident than in the center.
This fact contributes to lower the predicted $[<Mg/Fe>_{*}]$ in the outer regions, since the pop III yields predict a lower Mg/Fe ratio relative to the yields of normal massive stars. As a consequence, the $[<Mg/Fe>_{*}]$ gradient is shallower.

\section{Conclusions}
In this paper we have explored the effects of one or more early stellar 
generations of very massive stars (population III stars) on the chemical and 
photometric evolution of a typical elliptical galaxy of luminous mass 
$M_{lum}=10^{11}M_{\odot}$ and a dark matter halo ten times heavier.
To do that we have adopted models already tested on elliptical galaxies,
which reproduce the main properties of ellipticals (PM04). 
The adopted IMF has been the Salpeter (1955) one, with x=1.35 over the whole 
mass range.\par

We have included in the
chemical evolution models different nucleosynthesis results for very 
massive stars ($M >100 M_{\odot}$), taken both from recent and old 
literature.\par
Our results can be summarized as follows:
\begin{itemize}
\item Only one or two first stellar generations containing normal
stars plus very massive stars included in the best model of PM04 
(the galaxy forms 
by means of initial gas accretion), 
produce negligible effects on the subsequent chemical and photometric 
evolution, when either the yields for pair-creation SNe of HW02 or those 
of UN are adopted. Therefore, these models are acceptable. 

\item Population III stars (one or two stellar generations with 
nucleosynthesis either from UN or HW02)
in classic monolithic models for ellipticals (closed-box) produce  
larger, although still acceptable, $[<Mg/Fe>_{*}]$ ratios in the 
dominant stellar population 
relative to the standard case without population III pair-creation SNe.
However, the predicted integrated colors are too red. As a consequence these models should be rejected.

\item Models containing nucleosynthesis from Ober et al. (1983) for pair 
creation SNe, with one or two pop III stellar generations, produce quite 
different results relative to the other studied cases. In particular, in 
the framework of a monolithic closed-box model they produce a too high 
$[<Mg/Fe>_{*}]=4.66$,
completely at odd with what is observed ($[<Mg/Fe>_{*}]=0.2-0.3$, Thomas 
et al. 2002).

\item Models where the first stellar generations are made only of 
very massive stars followed by normal stellar generations generally 
produce lower $[<Mg/Fe>_{*}]$ ratios relative to the previous cases.
However, they all produce too high metallicities and too red integrated 
colors for ellipticals, therefore they should be ruled out.

\item It is worth noting that all these conclusions relative to abundance ratios would not change if the pop III stars were pre-galactic. The only difference would be found in the absolute abundances which would probably be more diluted.

\item We computed the metallicity distribution of stars at the present time 
for all the studied cases with pop III stars. We found that in 
most of the models the fraction of stars with metallicity lower than 
[Fe/H] =-2.5 is  negligible. This fact can create problems with the observed metallicity-luminosity relation (see Gibson 1996).
The lack of metal poor stars is due to the very fast chemical
enrichment produced by the pop III pair-creation SNe.
In particular, if we assume a threshold global metal content, for the formation of very massive stars, of $Z_{thr}=10^{-5} \cdot Z_{\odot}$, as recently suggested by Schneider et al. (2002), then the expected number of pair-creation SNe 
able to produce such a metallicity, in the framework of the infall model, is only between 10 and 15!

\item In summary, we conclude that if the population III stars formed for 
a very short time, their signature on the stellar and gaseous abundances are 
negligible and we cannot assess their existence nor disproved it.
If, on the other hand, they formed for a timescale not shorter than 0.1 Gyr 
then their effects are visible but produce results at variance with the main
observational properties of ellipticals.

\item Therefore, from the chemical and 
photometric point of view there is no need to invoke the existence of 
population III stars in ellipticals.
\end{itemize}

\section*{Acknowledgments} 

We thank C. Chiappini for many useful discussions.
The work was supported by MIUR under COFIN03 prot. 2003028039.


\begin{thebibliography}{}
\bibitem{}
\bibitem []{}Abel, T., Bryan, G., Norman, M.L., 2000, ApJ, 540,39 
\bibitem []{}Abel, T., Bryan, G., Norman, M.L., 2002, Science, 295, 93

\bibitem []{}Bond, H.E., 1981, AJ, 248, 606

\bibitem []{}Bromm, V., Coppi, P.S., Larson, R.B., 1999, ApJ, 527, L5
\bibitem []{}Bromm, V., Coppi, P.S., Larson, R.B., 2002, ApJ, 564, 23

\bibitem []{}Carr, B.J., 1987, Nature, 326, 829 
\bibitem []{}Carr, B.J., 1994, ARA\&A, 32, 531
\bibitem []{}Carr, B.J., Bond, J.R., Arnett, W.D., 1982, Proceedings of ESO
Workshop on ``The most Massive Stars'', eds. D'Odorico et al., Garching, Germany

\bibitem []{}Cayrel, R., 1986, A\&A, 168, 81
\bibitem []{}Cayrel, R., Depagne, E., Spite, M., Hill, V., Spite, F., et al. 2004, A\&A, 416, 1117
\bibitem []{}Cen, R., 2003a, ApJ, 591, L5
\bibitem []{}Cen, R., 2003b, ApJ, 591, 12
\bibitem []{}Chiappini, C., Romano, D., Matteucci,F., 2003, MNRAS, 339, 63

\bibitem []{}Chieffi, A., Dominguez, I., Limongi, M., Straniero, O., 2001, 
ApJ, 554, 1159

\bibitem []{}Chieffi, A., Limongi, M., 2004, ApJ, 608, 405 

\bibitem []{}Chiosi, C., Matteucci, F., 1983,
ESO Workshop on Primordial Helium, Garching, West Germany, February 2--3, 
1983, p. 77.

\bibitem []{}Christlieb, N., Bessel, M.S., Beers, T.C., Gustafsson, B., Korn, 
A., Barklem, P.S., Karlsson, T., Minuzo-Wiedner, M., 
Rossi, S., 2002, Nature, 419, 904

\bibitem []{}Christlieb, N., Gustafsson, B., Korn, A. J., Barklem, P. S., 
Beers, T. C., Bessell, M. S., Karlsson, T.,
Mizuno-Wiedner, M., 2004, ApJ, 603, 708

\bibitem []{}Ciardi, B., Ferrara, A., White, S.D.M., 2003, MNRAS, 344, L7

\bibitem []{}D'Antona, F., 1982, A\& A, 115,1
\bibitem[]{} De Donder, E., Vanbeveren, D., 2003, NewAstron., 8, 415

\bibitem []{}El Eid, M.F., Fricke, K.J., Ober, W.W., 1983, A\&A, 119, 54


\bibitem []{}Fran\c cois, P., Matteucci, F.Cayrel, R., Spite, M., Spite, F., 
Chiappini, C., 2004, A\&A, 421, 613


\bibitem[]{} Gibson, B.K., 1996, MNRAS, 278, 829

\bibitem []{}Greggio, L., \& Renzini, A. 1983, Mem SaIt, 54,311

\bibitem []{}Heger, A., Fryer, C.L., Woosley, S.E., Langer, N., Hartmann, 
D.H., 2003, ApJ, 591, 288

\bibitem []{}Heger, A., Woosley, S.E., 2002, ApJ, 567, 532 (HW02)

\bibitem []{}Iwamoto, N., Kajino, T., Mathews, G.J., Fujimoto, M.Y., Aoki, 
W.,  2004, ApJ, 602, 377

\bibitem[]{} Jimenez, R., Padoan, P., Matteucci, F., Heavens, A.F., 1998, 
MNRAS, 299, 123

\bibitem []{}Matteucci, F., \& Greggio, L., 1986, A\&A, 154, 279

\bibitem []{}Mehlert, D., Thomas, D., Saglia, R.P., Bender, R., $\&$ Wegner, G. 2003, A\&A, 407, 423

\bibitem []{}Nomoto, K., Hashimoto, M., Tsujimoto, T., Thielemann, F.K., Kishimoto, 
N., Kubo, Y., Nakasato, N., 1997, Nuclear Physics A, A621, 467

\bibitem []{}Ober, W.W., Fricke, K.J., El Eid, M.F., 1983, A\&A, 119, 61 (OFE83)


\bibitem []{}Pipino, A., Matteucci, F., 2004, MNRAS, 347, 968 (PM04)

\bibitem []{}Prantzos, N., 2003, A\&A, 404, 211
\bibitem []{}Salpeter, E.E., 1955, ApJ, 121, 161
\bibitem []{} Schaye, J., Aguirre, A., Kim, T.S., Theuns, T., Rauch, M., Sargent, W.L.W., 2003, ApJ, 318, 32
\bibitem[]{} Schneider, R., Ferrara, A., Natarayan, P., Omukai, K., 2002, ApJ, 571, 30
\bibitem []{}Siess, L., Livio, M., Lattanzio, J., 2002, ApJ, 570, 329
\bibitem []{}Songaila, A., 2001, Ap.J., 561, L153

\bibitem []{}Thielemann, F.K., Nomoto, K., Hashimoto, M. 1996, ApJ, 460, 408 

\bibitem []{}Thomas, D., Maraston, C., \& Bender, R., 2002, Ap\&SS, 281, 371

\bibitem []{}Umeda, H., Nomoto, K., 2002, ApJ, 565, 385 (UN)

\bibitem []{}van den Hoek, L.B., Groenewegen, M.A.T. 1997, A\&AS, 123, 305
\bibitem []{}Venkatesan, A., Schneider, R., Ferrara, A., 2004, 
MNRAS, 349, 43

\bibitem []{}Woosley, S.E., \& Weaver, T.A., 1995, ApJS, 101, 181 

\end{thebibliography}
\end{document}